\documentclass[aps,prl,twocolumn]{revtex4-1}

\usepackage{lipsum}
\usepackage{graphicx}
\usepackage{subfigure}
\usepackage{color}
\usepackage{amsmath}
\usepackage[linktocpage,colorlinks=true,linkcolor=blue,citecolor=blue,urlcolor=blue,breaklinks=true]{hyperref}
\usepackage{verbatim}
\usepackage{breakcites}
\usepackage{wrapfig}
\usepackage{soul}
\usepackage{bbold}
\usepackage{array,multirow}

\begin{document}

\widetext

\title{Confinement enhances the diversity of microbial flow fields}

\author{Rapha\"{e}l Jeanneret$^{1,3}$, Dmitri O. Pushkin$^2$, Marco Polin$^{3,4}$} 
\affiliation{$^1$IMEDEA, University of the Balearic Islands, Carrer de Miquel Marqu\`es, 21, 07190 Esporles, Spain\\
$^2$Mathematics Department, University of York, Heslington, York, YO10 5DD, UK\\
$^3$Physics Department, and $^4$Centre For Mechanochemical Cell Biology, University of Warwick, Gibbet Hill Road, Coventry CV4 7AL, United Kingdom}

\email[Correspondence: ]{M.Polin@warwick.ac.uk}

\begin{abstract}
Despite their importance in many biological, ecological and physical processes, microorganismal fluid flows under tight confinement have not been investigated experimentally. Strong screening of Stokelets in this geometry suggests that the flow fields of different microorganisms should be universally dominated by the 2D source dipole from the swimmer's finite-size body. Confinement therefore is poised to collapse differences across microorganisms, that are instead well-established in bulk.
 Here we combine experiments and theoretical modelling to show that, in general, this is not correct. Our results demonstrate that potentially minute details like microswimmers' spinning and the physical arrangement of the propulsion appendages have in fact a leading role in setting qualitative topological properties of the hydrodynamic flow fields of micro-swimmers under confinement. This is well captured by an effective 2D model, even under relatively weak confinement.
 These results imply that active confined hydrodynamics is much richer than in bulk, and depends in a subtle manner on size, shape and propulsion mechanisms of the active components. 
 \end{abstract}

\maketitle 

\begin{figure}
\centering
\includegraphics[width=1\linewidth]{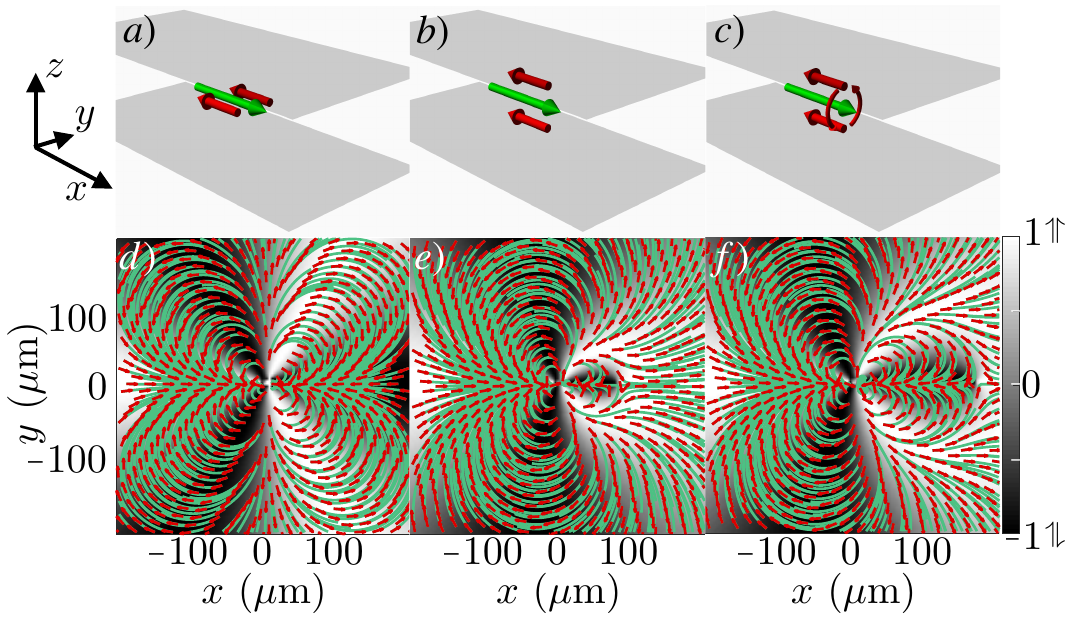}
\caption{The far-field signature of a force-free system depends on the relative z-position of the forces. Illustrated with a 3-forces system. a,d) The three forces are in a plane $\parallel$ to the plates: the resulting flow-field is quadrupolar with a $r^{-3}$ velocity decay. b,e) The forces are in a plane $\perp$ to the plates: the resulting flow-field is dipolar with a $r^{-2}$ velocity decay. c,f) Effect of spinning the plane of forces. The central quadrupolar region depends on both the radius of the orbit described by the forces and the on-axis separation between thrust and drag (red/green arrows respectively, in the first row). Gray-scale shows $\cos(\alpha)$, where $\alpha$ is the angle between the local velocity and the swimming axis $x$.}
\label{figure1}
\end{figure}

The way fluid is displaced around micro-swimmers is crucial to many biological, ecological and physical processes \cite{Lauga2009}. 
For instance, the uptake of nutrients and capture of small preys by micro-organisms depends directly on their flow fields \cite{Tam2011,Michelin2011,Dolger2017,Humphries2009,Jashnsaz2017, Mathijssen2018}; planktonic predators and preys detect each other mostly via fluid-mediated mechano-sensing \cite{Kiorboe1999,Jakobsen2006,Bruno2012,Kiorboe2013,Andersen2015}; and some species of protists can even relay information on potential nearby danger via hydrodynamic trigger waves \cite{Mathijssen2019}. The emergence of large-scale collective motion in microswimmers' suspensions is also set by the far-field symmetry of the fluid flows from individual active entities  \cite{Bricard2013,Stenhammar2017}. 
Microscopic fluid disturbances naturally fall into the inertialess regime (low Reynolds number), and are governed by  the Stokes equations. In this regime, flows can be decomposed and expanded in terms of singularity solutions, or multipoles \cite{KimKarrila}. 
In unbounded fluids, the almost neutrally buoyant swimming microorganisms are  generally modelled either as basic force dipoles (stresslets), or with spatially extended dipole variants like the 3-forces model introduced for the microalga \textit{Chlamydomonas reinhardti} (CR) \cite{Drescher2010} (see Fig.~\ref{figure1}a). The resultant flow fields decay as $\sim r^{-2}$ \cite{Drescher2010, Drescher2011}, and the sign of the effective force dipole divides micro-swimmers into two large classes: pushers (e.g. bacteria, pushing fluid with their rear-mounted flagella) and pullers (e.g. CR, pulling the fluid with front-mounted cilia). This division appears to be very important in setting macroscopic properties of active fluids, from flow instabilities to bulk rheology \cite{Lauga2009, Rafai2010, Lopez2015}.
Biological and artificial active particles, however, are often confined within boundaries, either as a consequence of their natural habitat \cite{Foissner1998, Or2007, Kantsler2013,Elgeti2015}, or for technological purposes \cite{Denissenko2012}, or simply to facilitate experiments \cite{Guasto2010,Pepper2010}. 
In this context, theory has predicted that the bulk flow picture should be critically modified by the boundaries \cite{Brotto2013, Delfau2016}. Here we provide a systematic experimental test of these confinement-induced changes in microbial flows.

Important differences are expected in the multipolar expansions of flows from microorganisms between the bulk and confined cases.
A point-force confined between two parallel no-slip walls creates, in the far field, a fluid disturbance akin to a 2D source dipole with a velocity decay $\sim r^{-2}$ \cite{Liron1976}. As noted in \cite{Brotto2013, Delfau2016}, a force-dipole between two plates should then produce a far-field flow decaying as $\sim r^{-3}$, much faster than in bulk. At the same time, whether driven by flows, sedimenting, or self-propelled, a \textit{finite-sized} particle moving at a velocity different from the background fluid must induce a source-dipole perturbation along the direction of motion to fulfil mass conservation. In a quasi-2D Hele-Shaw configuration, when the swimmer's size $d$ is comparable to the confinement length $H$, this singularity decays as $\sim r^{-2}$ \cite{Diamant2009,Beatus2012,Janssen2012,Desreumaux2013,Brotto2013}, and should therefore dominate the multipolar expansion regardless of the arrangement of propulsive and drag forces.
Consequently, confinement should substitute the bulk division between pushers and pullers with a single class of micro-swimmers whose far-field hydrodynamic interactions are universally mediated by 2D source-dipoles, although numerical studies suggest that very near-field details might also be important \cite{Delfau2016}. 
These predictions stand in stark contrast with a fundamental lack of systematic experimental investigations to test and substantiate the theoretical picture (but see \cite{Kruger2016,Thutupalli2018} for collective effects in confined active droplets).

In this letter, we combine systematic experiments with modelling to show that, within the experimentally accessible range, confinement does not lead to a universal collapse of  microbial flows. Instead, we observe strong qualitative differences resulting from details in the geometry and propulsion of different microbial species.  
Intuitively, these can be understood to arise from the dependence of wall-induced screening of forces on the forces' position across the sample cell, with the net result to multiply the variety of microbial flow fields with respect to the bulk case. 
Despite their sensitivity to the spatial structure of the micro-swimmer and the level of confinement, the experimental flow fields can be modelled  accurately within a  2D thin-film approximation even under relatively weak confinement.

To clarify the effect of confinement, we begin with a simple example. Liron and Mochon \cite{Liron1976} showed that a point-force $\mathbf{F}_{\parallel}=(F_x,F_y,0)$ located at $(x=0,y=0,z=h)$ within a Hele-Shaw cell of thickness $H$ in the $z$-direction, generates a far-field flow given by 
\begin{equation}
\mathbf{u}_{LM,\parallel}(\mathbf{r},z)=f(z,h,H)(\mathbb{1}/r^2-2\mathbf{r}\mathbf{r}/r^4)\cdot\mathbf{F}_{\parallel},
\end{equation}
where $\mathbf{r}=(x,y)$ and $f(z,h,H)=-3H/(2\pi\mu)[z/H-(z/H)^2][h/H-(h/H)^2]$. This field is equivalent to a 2D source dipole whose strength $f(z,h,H)$ depends quadratically on the vertical position $h$ of the Stokeslet, with a maximum in the mid-plane ($h=H/2$). 
Ignoring temporarily finite-size effects for real microswimmers, this $h$-dependence immediately implies that the flow-field created by a force-free swimmer should be qualitatively very sensitive to the spatial arrangement of forces along the $z$-direction, because the relative effect of these forces on the fluid can be very different.  
To illustrate this point, let us consider the 3-Stokeslets model for CR \cite{Drescher2010}, where a single force on the fluid representing the cell body motion ($\mathbf{F}_{\parallel}$; Fig.~\ref{figure1}a-c green arrow), is balanced by a pair of forces representing the two front flagella ($-\mathbf{F}_{\parallel}/2$; Fig.~\ref{figure1}a-c red arrows). When the forces are parallel to the $xy$-plane, the far-field has indeed a force-dipole symmetry, with a $\sim r^{-3}$ decay as predicted in \cite{Brotto2013} (Fig.~\ref{figure1}a,d). However, when the forces lay on a plane {\it perpendicular} to the $xy$-plane, the far-field has a source dipole symmetry with a slower $\sim r^{-2}$ decay (Fig.~\ref{figure1}b,e).
The size of the force-dipole-like recirculation region close to the front of the swimmer (Fig.~\ref{figure1}e) depends strongly on both on-axis distance between thrust and drag forces, and  the separation between the putative flagellar forces. 
For a swimmer that spins as it swims, as for CR, the topology of the flow-field will then oscillate periodically as a function of the rotation of the flagellar plane (see Movie~S1 \cite{supplementary}), and not just as a function of the phase in the beating cycle \cite{Guasto2010,Klindt2015}. In this case, the rotation-averaged flow always retains a source dipole far-field  symmetry (Fig.~\ref{figure1}c,f), while the extent of the near-field recirculation depends on the separation between the pair of thrust forces (Fig.~S1 \cite{supplementary}). 
Overall, these arguments suggest that the effective 2D representation of the far-flow field of a confined force-free micro-swimmer should be guessed with care, as the induced flow has a strong qualitative dependence on the spatial arrangement of the swimmer's propulsion and drag forces.
This sensitivity is in sharp contrast with the equivalent case in bulk. It can be understood intuitively as a result of the $h$-dependence of the function $f(z,h,H)$, which implies that a Stokeslet in the mid-plane produces a $z$-averaged far-flow field stronger than one outside it. This consequence of confinement appears to have been largely overlooked, but it could be put to good use to build artificial active systems with {\it in situ} tuneable hydrodynamic interactions, for instance by modulating the arrangement or orientation of active particles across the Hele-Shaw cell through the application of external fields. Such systems should display a rich set of collective dynamic phenomena.
 
\begin{figure*}
\centering
\includegraphics[width=0.95\linewidth]{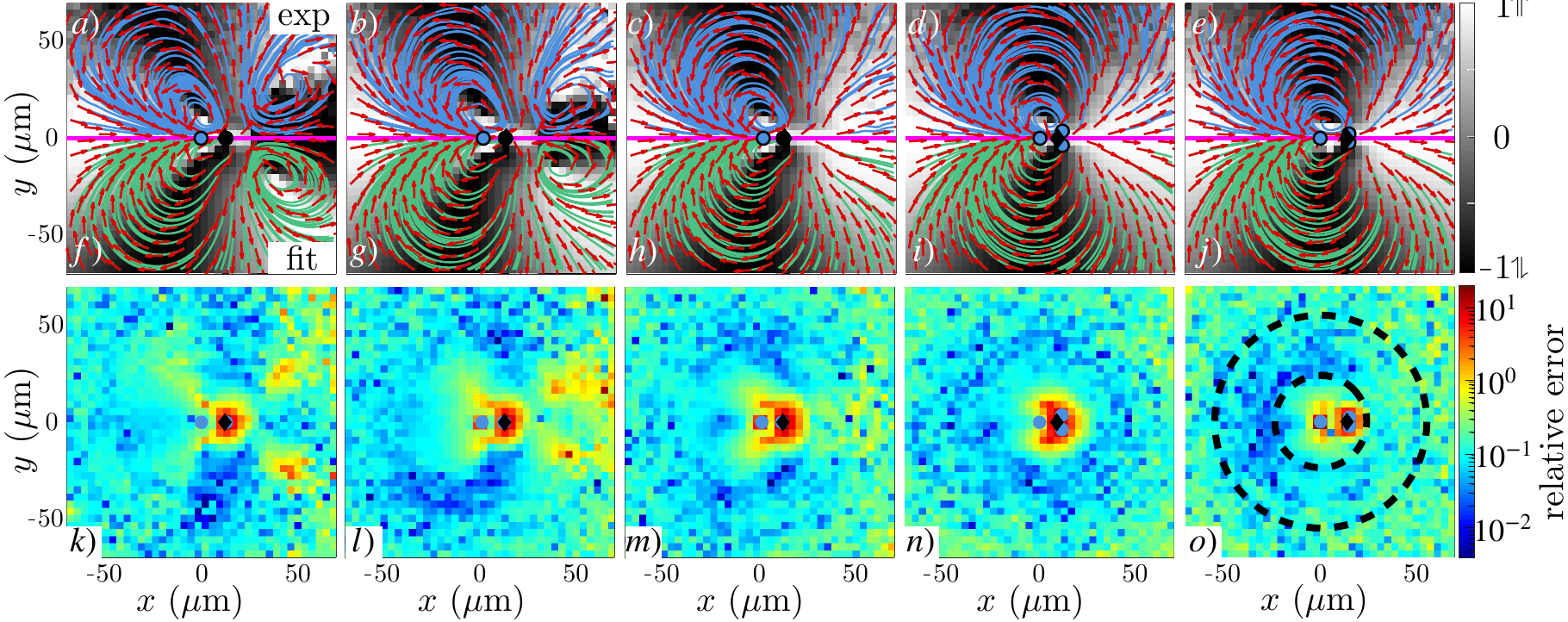}
\caption{CR flow fields. a-e) Experimental flow-fields for CR for: a) $H=60\,\mu$m, b) $H=43\,\mu$m, c) $H=29\,\mu$m, d) $H=21.5\,\mu$m, e) $H=14\,\mu$m. f-j) Corresponding flow-fields obtained from best fits to the effective 2D model. Blue dots: Stokeslets' location; black dot: source dipole location. Gray scale follows convention from Fig.~\ref{figure1}. All panels show streamlines (solid lines) and local velocity vector fields (red arrows). k-o) Magnitudes of the difference between experimental and fitted flows, normalised by the experimental magnitude. Dashed circles in  o) indicate the region of  the experimental flows used to determine best fits.}
\label{figure2}
\end{figure*}

With this in mind we now turn to the experiments, measuring  flow-fields under controlled confinement for both a puller-like and a pusher-like swimmer, respectively CR and the dinoflagellate \textit{Oxyrrhis marina} (OM). These organisms have similarly shaped prolate cell bodies (Fig.~S2a,b \cite{supplementary}), with diameters $d_{\rm CR}\lesssim10\,\mu$m and $d_{\rm OM}\lesssim 20\,\mu$m respectively. The former propels with a characteristic breaststroke beating of its pair of front-mounted flagella $\sim12\,\mu$m-long; the latter employs a $\sim30\,\mu$m-long back-mounted flagellum which propagates bending waves (Movies~S2,3; Fig.~S2a,b \cite{supplementary}). Both species spin as they swim. The microorganisms were grown following \cite{Mathijssen2018}, and then loaded in microfluidic chambers of uniform thickness $H$ ($14\,\mu\text{m}\leq H\leq 60\,\mu$m) previously passivated with a $0.5\%\,$w/v Pluronic F-127 solution. Tracking of $1\,\mu$m polystyrene tracer particles (Polysciences, USA) in the reference frame centred on the microorganism and oriented along its swimming direction, was done through a $20\times$ NA 0.40 objective (Nikon, Japan) at $50\,$fps. This allowed us to reconstruct the induced flow field averaged over the spinning and beating cycles, and across the $10\,\mu$m focal volume centred in the middle of the chamber \cite{Drescher2010,supplementary}.
Figure~\ref{figure2}a-e shows the flow fields for CR at five decreasing chamber thicknesses (experimental/numerical flow fields in blue/green throughout).
Under weak confinement ($H=60,43\,\mu$m; Fig.~\ref{figure2}a,b) the flows present a characteristic puller-like symmetry, with a stagnation point $\sim25\,\mu$m in front of the cell. Although both features are typical of bulk flows \cite{Drescher2010}, the bulk solution yields in fact a poor quantitative agreement, even for $H=60\,\mu$m (Fig.~S3 \cite{supplementary}).
As the channel thickness decreases further ($H=29,21.5,14\,\mu$m; Fig.~\ref{figure2}c,d,e) the flow-field develops clearly the structure of a source dipole. The velocity decays as $\sim r^{-2}$, confirming that this is a two-dimensional source dipole (Fig.~S4 \cite{supplementary}). At the same time, close to the cell  the flow presents some differences from a pure 2D source dipole, with a slight front-back asymmetry and side vortices (Fig.~\ref{figure2}c-e).

\begin{figure}
\centering
\includegraphics[width=0.85\linewidth]{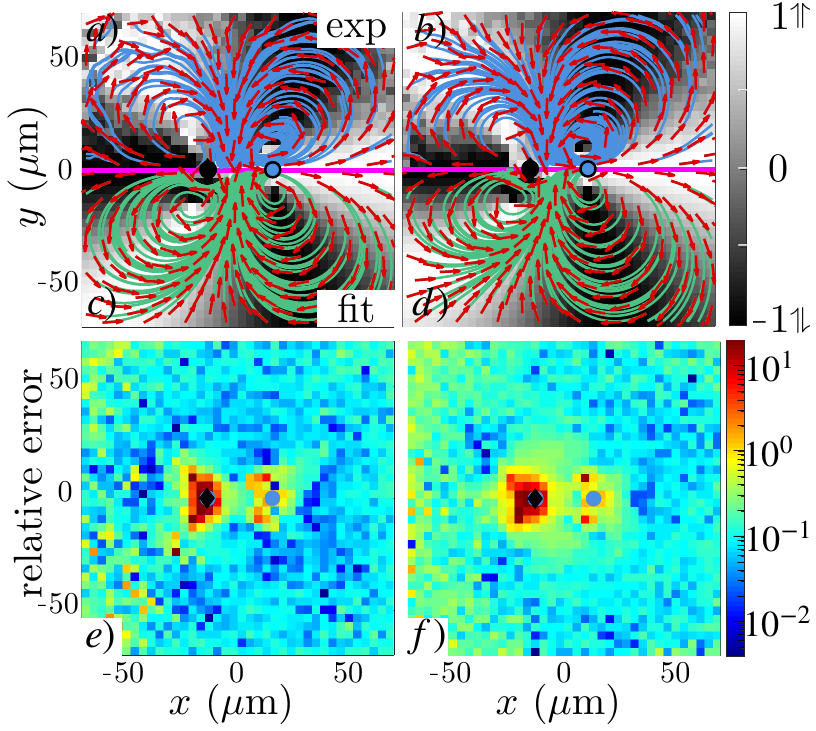}
\caption{OM flow fields.
Experimental flow-fields for OM for: a) $H=21.5\,\mu$m, b) $H=29\,\mu$m. c,d) Corresponding flow-fields obtained from best fits to the effective 2D model. Blue dots: Stokeslets' location; black dot: source dipole location. Gray scale follows convention from Fig.~\ref{figure1}. All panels show streamlines (solid lines) and local velocity vector fields (red arrows). e,f) Magnitudes of the difference between experimental and fitted flows, normalised by the experimental magnitude.}
\label{figure3}
\end{figure}

Figure~\ref{figure3}a,b, however, shows that the flows generated by OM in strong confinement ($H=21.5,29\,\mu$m) are qualitatively different from those observed for CR. Within our experimentally accessible range, corresponding to a $\sim30\times$ velocity decay for both species (Figs.~S4,5 \cite{supplementary}), OM flows display a front-back asymmetric force dipole field instead of CR's source dipole. 
This striking difference in flow structure confirms that confinement does not reduce all microbial flows to a unique type, but rather makes them very sensitive to precise details of a swimmer's geometry beyond its finite size body.

To rationalise the measured flows, it is instructive to start first with a simple superposition of the far-field point force solutions from Liron and Mochon \cite{Liron1976}. Fitting the average flows of a spinning 3-forces model for CR (Fig.~\ref{figure1}c) and off-centre 2-forces one for OM (Fig.~S2 \cite{supplementary}) to the experimental flow-fields in $H=21.5\,\mu$m, reveals clearly that both models lack an extra 2D source dipole (Fig.~S6 \cite{supplementary}). This would naturally arise from the cells' finite-size bodies \cite{Brotto2013}, and suggests
to turn to a conceptually simpler 2D approach, in the spirit of the general treatment of Hele-Shaw flows \cite{Pushkin2016}.
The microorganisms, centred in the field of view and swimming along the positive $x$-direction, are modelled by a set of point forces representing drag (strength $F_{\rm S}$, position $(x_0,0)$) and thrust (CR: two point forces of strength $-F_{\rm S}/2$ at $(x_1,\pm y_1)$; OM: one point force of strength $-F_{\rm S}$ at $(x_1,0)$). Each force is along the $x$ axis, and generates a flow given by the Green's function for the effective 2D Stokes equations \cite{Pushkin2016, supplementary}. For no-slip boundaries, the functional shape of this flow depends on a single lengthscale $\lambda=H/\sqrt{12}$, fixed here by the measured sample thicknesses. The forces are then supplemented by a 2D source dipole $\mathbf{u}_{d}=-I_{\rm d}(\mathbf{e}_x/r^2-2\mathbf{r}x/r^4)/2\pi$ at position $(x_{\rm d},0)$.
It represents both the effect of  finite-size body, and the unequal screening of drag and thrust forces by the walls which is connected to the organism's shape and spinning (see Fig.~\ref{figure1}c,f). 
The best fits to the experimental data are shown in Fig.~\ref{figure2}f-j and Fig.~\ref{figure3}c-d. They were obtained through a systematic sweep in the space of initial parameter values \cite{supplementary} searching optimal fits to an annular region between $25\,\mu$m and $60\,\mu$m around the swimmer (Fig.~\ref{figure2}o, dashed lines). Within this region, for each one of the $3.7\,\mu\text{m}\times3.7\,\mu$m spatial bins, we collected at least $7\times10^3$ independent measurements for CR (370 for OM) for each $H$ value (Table~S2 \cite{supplementary}). The fits agree very well with the experiments, with typical relative errors of $15\%$ (Fig.~\ref{figure2}k-o, Fig.~\ref{figure3}e,f; Table~S3 \cite{supplementary}). The model captures also subtle details of the experimental flows. For CR these include: the deformation of the streamlines in front of the organism; the approximate location of the side-vortices under strong confinement; the location of the stagnation point under weak confinement. For OM the model reproduces well the front-back asymmetry of the experimental force-dipole-like field.

\begin{figure}[b]
\centering
\includegraphics[width=0.95\linewidth]{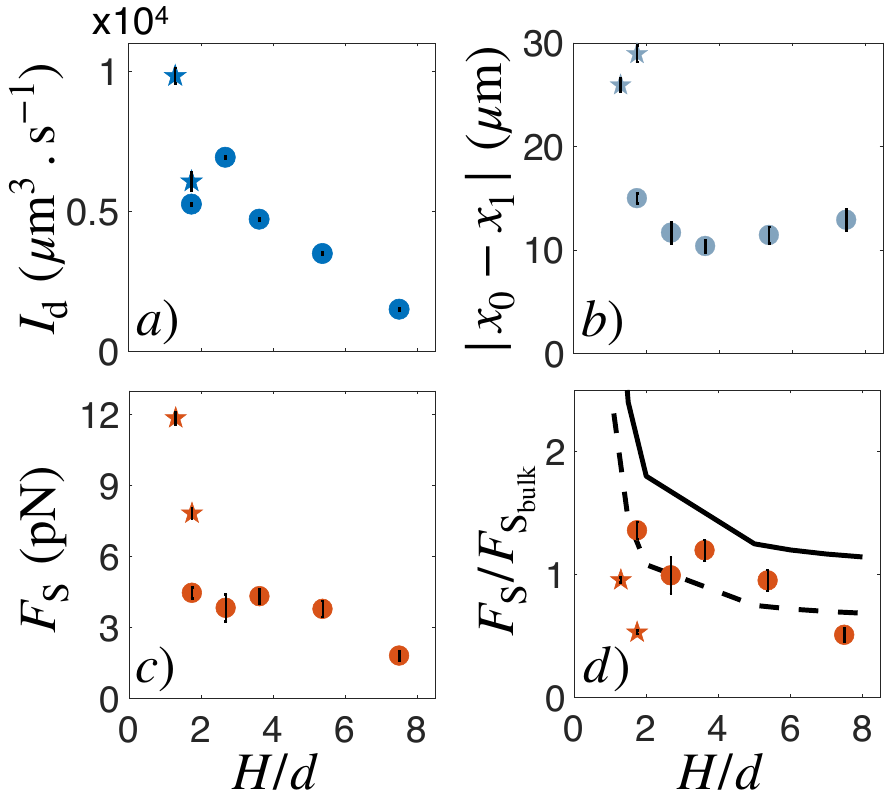}
\caption{Dependence of a) the dipole strength $I_{\rm d}$, b) on-axis force separation $|x_0-x_1|$, c) 2D Stokeslet strength $F_{\rm S}$ as a function of normalised sample thickness $H/d$ for all the micro-organisms studied. OM: $\star$; CR: $\circ$. d) $F_{\rm S}$ normalised by the bulk drag  $F_{\rm S_{\text{bulk}}}$ on CM and OM cell bodies modelled as prolate ellipsoids moving at the measured $H$-dependent speeds. Solid line: prediction for a sphere of diameter $d$ moving at the centre of the Hele-Shaw cell \cite{Ganatos1980}; dashed line is the solid line diminished by $40\%$. Error bars throughout represent fit uncertainties to average experimental flow fields.}
\label{figure4}
\end{figure}

Figure~\ref{figure4} shows the dependence of the main fitting parameters on the reduced sample thickness $H/d_{(\text{CR,OM})}$. The others (CR:$(x_d,x_0,y_1)$; OM:$(x_d,x_0)$), which encode the spatial structure of the organisms, are consistent across $H$ values  (see Fig.~S7 in \cite{supplementary}). The error bars represent fit uncertainties to the average experimental flow fields.
As expected for weakening confinement, the dipole strength $I_{\rm d}$ decreases steadily as $H/d$ increases (Fig.~\ref{figure4}a), along what appears to be a single curve for both microorganisms. 
The 2D Stokeslet strength $F_{\rm S}$, responsible for the thrust, is larger for OM than for CR ($\sim10\,$pN vs. $4.2\pm0.3\,$pN respectively, Fig.~\ref{figure4}c), mirroring differences in size and speed which lead to higher drag for OM than for CR. Values for CR are in line with previous estimates \cite{Klindt2015} and largely independent of confinement, although decrease noticeably for $H=60\,\mu$m. 
The fitted propulsive forces $F_{\rm S}$ can be normalised by the bulk drag for prolate ellipsoids mimicking the swimmers' bodies, and translating along the major axis at the measured $H$-dependent swimming speed. This provides an estimate of the increase in cell-body drag within the Hele-Shaw cells which can be compared with the values predicted for a sphere of radius equal to the cells' semi-minor axis (Fig.~\ref{figure4}d, solid line)\cite{Ganatos1980,supplementary}. The latter appears to systematically overestimate the experimental drag estimate by  $\sim40\%$ (Fig.~\ref{figure4}d, dashed line), suggesting that, despite the excellent agreement between the flow fields, $F_{\rm S}$ might underestimate the full propulsive force of the confined microorganisms. This possibly results from momentum transfer to the surrounding walls in the immediate vicinity  of the microorganism.
 Finally, Fig.~\ref{figure4}b shows that the fits return an on-axis separation between drag and thrust forces which is both of the correct magnitude and largely independent of the sample thickness $H$. The large separation for OM is ultimately at the origin of the large asymmetric force-dipole-like flow  observed for this organism (Fig.~\ref{figure3}a,b). By artificially reducing this parameter, the flow acquires a source-dipole structure (Fig.~S8 \cite{supplementary}).

In this letter we presented what is to the best of our knowledge the first systematic experimental study of the effect of confinement on micro-swimmers' hydrodynamics.
In line with previous studies \cite{Brotto2013, Delfau2016}, the finite size body and common spinning motion of microorganisms are expected, and observed, to produce a far field 2D source dipole. However, experiments with OM highlight that the spatial structure of a microorganism can easily push this far field to distances where flows are, for practical purposes, negligible. In the present case, this leads to strong differences in the topology of OM and CR flows, as a result of the on-axis separation between OM's propulsion and drag forces. 
We expect that these qualitative differences will influence both the biology (e.g. feeding currents) and the physics (e.g. collective behaviour) of microorganisms in confinement.
Despite qualitative differences, the flows of both micro-swimmers can be accurately described by considering just a 2D source dipole and a force-free combination of Stokeslets \cite{Pushkin2016}. The latter should reflect the specific arrangement of swimming appendages for each microorganism. We hope that our work will inspire future investigations on the great diversity of fluid flows in confined active matter.

\begin{acknowledgments}
We thank Enkeleida Lushi for insightful discussions and encouragement. This work was partly supported by a Margalida Comas Fellowship (PD/007/2016) (RJ).
\end{acknowledgments}

\bibliography{Biblio_flowfield}

\newpage

\section{Confinement enhances the diversity of microbial flow fields}

\renewcommand{\thetable}{S\arabic{table}}

\setcounter{table}{0}
\renewcommand{\arraystretch}{1.5}

\begin{table*}
	\centering
	\caption{Swimmers characteristics.}
\begin{tabular}{|c|c|c|c|}\hline
\multirow{1}{*}{Organism} & Swimming speed $({\rm \mu m.s^{-1}})$ & Major axis length $({\rm \mu m})$ & Minor axis length $({\rm \mu m})$ \\  \hline
\multirow{5}{*}{CR} & $H=14 {\rm \mu m}\Rightarrow  V=41.3\pm13.0$ &  &  \\\cline{2-2}
    & $H=21.5 {\rm \mu m}\Rightarrow  V=48.5\pm15.0$ &  &  \\ \cline{2-2}
    & $H=29 {\rm \mu m}\Rightarrow  V=45.4\pm17.5$ & $10.1\pm1.7$ & $8.0\pm1.7$ \\ \cline{2-2}
		 & $H=43 {\rm \mu m}\Rightarrow  V=50\pm18.0$ &  &  \\ \cline{2-2}
		 & $H=60 {\rm \mu m}\Rightarrow  V=44.6\pm18.3$ &  &  \\\hline
\multirow{2}{*}{OM} & $H=21.5 {\rm \mu m}\Rightarrow  V=73.6\pm15.6$ & $22.9\pm3.4$ & $16.6\pm1.8$\\ \cline{2-2}
    & $H=29 {\rm \mu m}\Rightarrow  V=87.5\pm23.9$ &  & \\
\hline
\end{tabular}
\label{TableS1}
\end{table*}

\section{Description of the experiments}
The measurement of the flow-fields was performed as previously done in [1] (Drescher et al PRL 2010) by using Particle Tracking Velocimetry (PTV) to reconstruct the motion of $1\,\mu$m-polystyrene particles (Polysciences, cat. no. 19819-1) in the frame of reference centred on the microorganism and with the $x$-axis oriented along the instantaneous swimming direction. This registers all the tracer displacements as if they were measured in the laboratory frame with the swimmer passing by the origin with its velocity oriented along the positive $x$-axis.
The mixed cells/colloids suspension was loaded in microfluidic Hele-Shaw channels of well-controlled thickness which were then sealed with vaseline to prevent flows from evaporation. Before loading the suspension, we left a $0.5\%$ w/v Pluronic F-127 solution inside the chamber for at least 30 minutes in order to passivate the surfaces of the microfluidic chips and limit the sticking of the particles. 
To limit the amount of data acquired, we restrained ourself to the measurement of flow-fields averaged over the strokes and spinning of the organisms, which was done by recording at $50$ fps. In addition, in order to optimise the statistics per frame while keeping the cells concentration low enough we recorded at 20x magnification (under Phase Contrast illumination) with a 1 Megapixels camera (model: Pike F-100B, AVT). This gives a square field of view of width $370 {\rm \mu m}$ in which we had on average $\sim 2-3$ cells. This setting implies also that the depth of focus is relatively large ($\lesssim 10 {\rm \mu m}$), leading to measured flow-fields vertically averaged over this length. Finally we always focused in the middle of the chambers to ensure a symmetric situation with respect to the two confining walls.

 We have measured the flow-fields of CR (CC-125) and OM (CCAP-1133/5) 
 in chamber thickness $H=14, 21.5, 29, 43, 60\,\mu$m and $H=21.5, 29\,\mu$m respectively (Figs.2,3 main text). The average speed and size of the swimmers are collated in Table \ref{TableS1}, together with error bars representing the standard deviation of the respective distributions across the population (and not the errors of the mean). For each $H$ value, we recorded at least 1000 individuals for CR and at least 175 individuals for OM.

\section{Effective 2D model}

We modelled the flow produced by the micro-organisms as a multipolar expansion by considering a pure 2D source dipole that represents the effect of swimmers' finite size and a set of 2D Stokeslets that represents propulsion. The source dipole is located at $(x_{\rm d},0)$ and is oriented along the direction of motion $x$: 

\begin{align}
\label{solution_Stokeslet}
u_{{\rm d},x}(x,y)&=\frac{I_d}{2\pi}\frac{(x-x_{\rm d})^2-y^2}{r^4},\\
u_{{\rm d},y}(x,y)&=\frac{I_d}{2\pi}\frac{2(x-x_{\rm d})y}{r^4},
\end{align}

The 2D Stokeslet is given by \cite{Pushkin2016}: 

\begin{align}
\label{solution_Stokeslet}
\mathbf{u}_{\mathbf{f}}(\mathbf{r},\mathbf{F})&=\mathbf{G}_{\mathbf{f}}.\frac{\mathbf{F}}{2\pi \eta H},
\\
\mathbf{G}_{\mathbf{f}}&= f_1(r/\lambda)\mathbf{I}+f_2(r/\lambda)\hat{\mathbf{r}}\hat{\mathbf{r}},
\\
f_1(w)&=K_0(w)-(w^{-2}+w^{-1}K_0^{\prime}(w)),
\\
&=K_0(w)-w^{-2}+w^{-1}K_1(w),
\\
f_2(w)&=2w^{-2}-w(w^{-1}K_0^{\prime}(w))^{\prime},
\\
&=2w^{-2}-2w^{-1}K_1(w)-K_0(w),
\end{align}
where $\mathbf{F}$ is the point force, $\eta$ is the dynamic viscosity of the fluid and $K_{0,1}$ is the modified Bessel function of the second kind of zero-th (resp. first) order. In this formula the point-force is located at the origin. For no-slip surfaces, the length $\lambda$ is related to the thickness of the sample cell $H$ by $\lambda=H/\sqrt{12}$ (notice that there is a typo in \cite{Pushkin2016}, which states instead that $\lambda=H$).

To model CR we consider three of those Stokeslets, one at $(x_{0},0)$ oriented along the direction of motion which models the drag on the cell body (strength $F_{\rm S}$), and two others located at $(x_{1},y_{1})$ and $(x_{1},-y_{1})$, oriented in the opposite direction, which represent the thrust from the flagella. Each of those thrust forces has a strength $F_{\rm S}/2$. 

To model OM we consider only two of those Stokeslets, one at $(x_{0},0)$ oriented along the direction of motion which models the drag on the cell body (strength $F_{\rm S}$), and an other located at $(x_{1},0)$, oriented in the opposite direction, which represents the thrust from the flagellum (strength $F_{\rm S}$). 

\begin{table*}
	\centering
	\caption{Statistics of the fitting procedure.}
	\begin{tabular}{|c|c|c|c|c|} \hline
		Organism and $H$ &Avg. reads/bin ($r_{\text{in}}<r<r_{\text{out}}$)& Total no. of fits &  No. of occurences of selected fit & Rel. error of fit selected ($\times 10^{-3}$)\\ \hline
		CR $14 {\rm \mu m}$ &$8.4\times10^3$& $1800$  & $202$ & $7.5$ \\ \hline
		CR $21.5 {\rm \mu m}$ &$7.2\times10^3$& $1800$  & $125$ & $8.9$ \\ \hline
		CR $29 {\rm \mu m}$ &$2.0\times10^4$& $1800$  & $44$ & $8.4$ \\ \hline
		CR $43 {\rm \mu m}$ &$1.8\times10^4$& $1800$  & $397$ & $18.6$ \\ \hline
		CR $60 {\rm \mu m}$ &$4.7\times10^4$& $1800$  & $321$ & $29.6$ \\ \hline
		OM $21.5 {\rm \mu m}$ &$923$& $600$  & $235$ & $45.9$ \\ \hline
		OM $29 {\rm \mu m}$ &$370$& $600$  & $424$ & $72.2$ \\ \hline		
	\end{tabular}	
	\label{TableS2}
\end{table*}

\begin{table*}
	\centering
	\caption{Relative errors of the fits: median values.}
	\begin{tabular}{|c|c|c|c|} \hline
		Organism and thickness & 2D thin film &  Liron\&Mochon Stokeslets model & Bulk Stokeslets model\\ \hline
		CR $14 {\rm \mu m}$ & $11\%$  & $$ & $$ \\ \hline
		CR $21.5 {\rm \mu m}$ & $9.5\%$  & $73\%$ (Fig.~\ref{figureS6})& $$ \\ \hline
		CR $29 {\rm \mu m}$ & $10\%$  & $$ & $$ \\ \hline
		CR $43 {\rm \mu m}$ & $15\%$  & $$ & $$ \\ \hline
		CR $60 {\rm \mu m}$ & $20\%$  & $$ & $52\%$ (Fig.~\ref{figureS3}) \\ \hline
		OM $21.5 {\rm \mu m}$ & $18\%$  & $53\%$ (Fig.~\ref{figureS6})& $$ \\ \hline
		OM $29 {\rm \mu m}$ & $30\%$  & $$ & $$ \\ \hline		
	\end{tabular}	
	\label{TableS3}
\end{table*}

\section{Procedure for the fitting of the flow-fields}

We used the model described in the previous section to fit our experimental flow-fields. Consequently we have 6 free-fitting parameters for CR $(I_{\rm d}, x_{\rm d}, F_{\rm S}, x_0, x_1, y_1)$ and 5 for OM $(I_{\rm d}, x_{\rm d}, F_{\rm S}, x_0, x_1)$. Because of this large number, the fitting procedure is very sensitive to the initial guess values imposed on these parameters. There are many local minima in this high-dimensional space. Then, in order to select the best possible fit amongst these local minima, we have systematically performed a sweep on the initial guess values. The final choice of the best fitting parameters was then based on the relative error of the fit, the probability of its occurence and the physical soundness of the parameters value (see Table~\ref{TableS2}). We used a non-linear least-square approach using the Matlab function fminsearch and minimizing the relative error defined as $\epsilon(\{x_i,y_j\})=\sum_{i,j} (\mathbf{u}_{\rm exp}(x_i,y_j)-\mathbf{u}_{\rm fit}(x_i,y_j))^2/\sum_{i,j} \mathbf{u}_{\rm exp}^2(x_i,y_j)$. Error bars on the fitting parameters have been estimated by computing the variance-covariance matrix obtained with a direct least-square fitting of the 2D flows using the Matalb function lsqcurve and setting the initial guess values of the parameters as the one given by the best fit from the first approach. Finally the fitting has been performed within a ring defined by the two radii $r_{\rm in}=25 {\rm \mu m}$ and $r_{\rm out}=60 {\rm \mu m}$ from the center of the organisms. The value of $r_{\rm out}$ corresponds to the limit where the signal-to-noise ratio becomes too low, while $r_{\rm in}$ has been chosen to limit the influence of the very details of the near-field flows.

\bibliography{Biblio_flowfield}

\section{Supplementary figures}

\renewcommand{\thefigure}{S\arabic{figure}}

\setcounter{figure}{0}

\begin{wrapfigure}{r}{8cm}
\centering
\includegraphics[width=\linewidth]{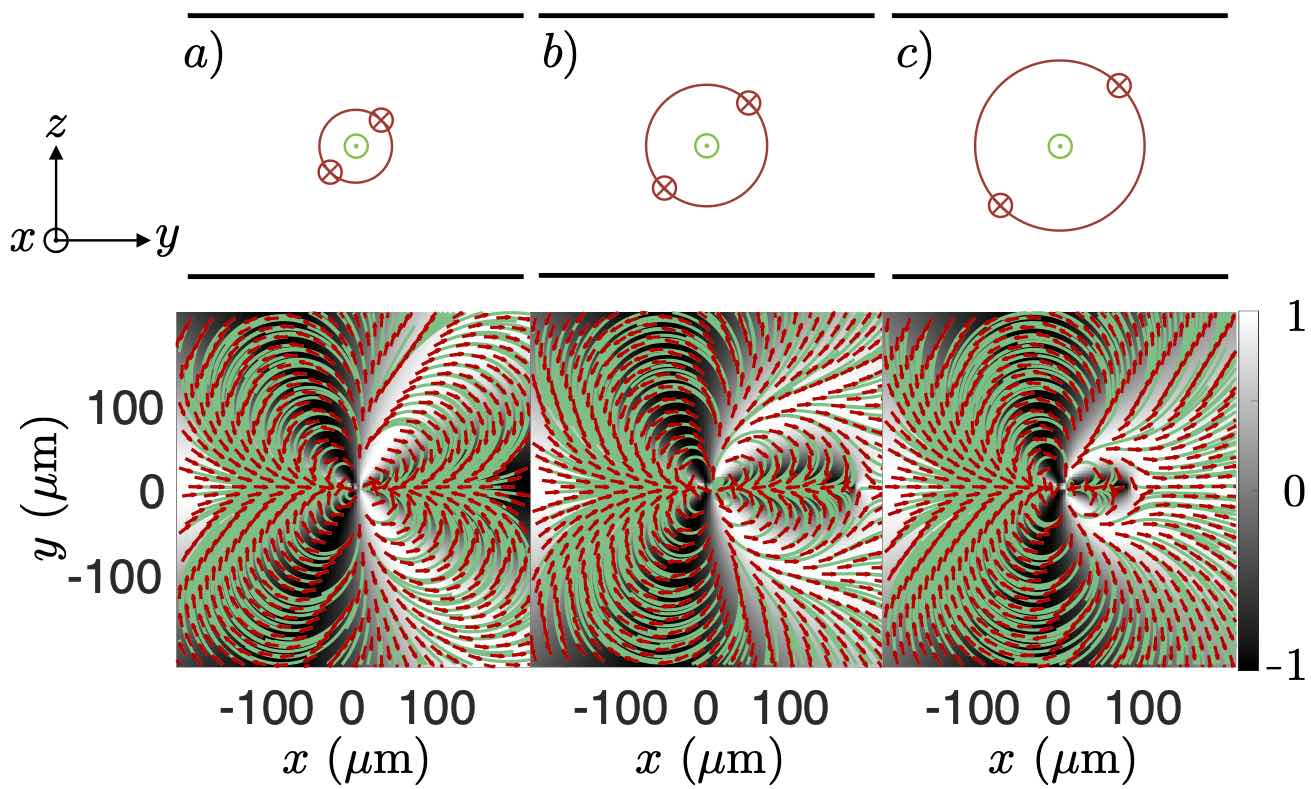}
\caption{A spinning 3-forces swimmer always generates (on average) a dipolar far-field with a quadrupolar recirculation zone in the center. The extent of this region directly depends on the radius of the orbit described by the thrust forces. We show here the flow-fields obtained in $H=21.5 {\rm \mu m}$ for a 3-forces swimmer having the drag force (green) in the mid-plane and the thrust forces (red) $10 {\rm \mu m}$ in front and describing a circular orbit with radius: a) $3 {\rm \mu m}$, b) $5 {\rm \mu m}$, c) $7 {\rm \mu m}$.}
\label{figureS1}
\end{wrapfigure}

\begin{wrapfigure}{r}{8cm}
\centering
\includegraphics[width=0.8\linewidth]{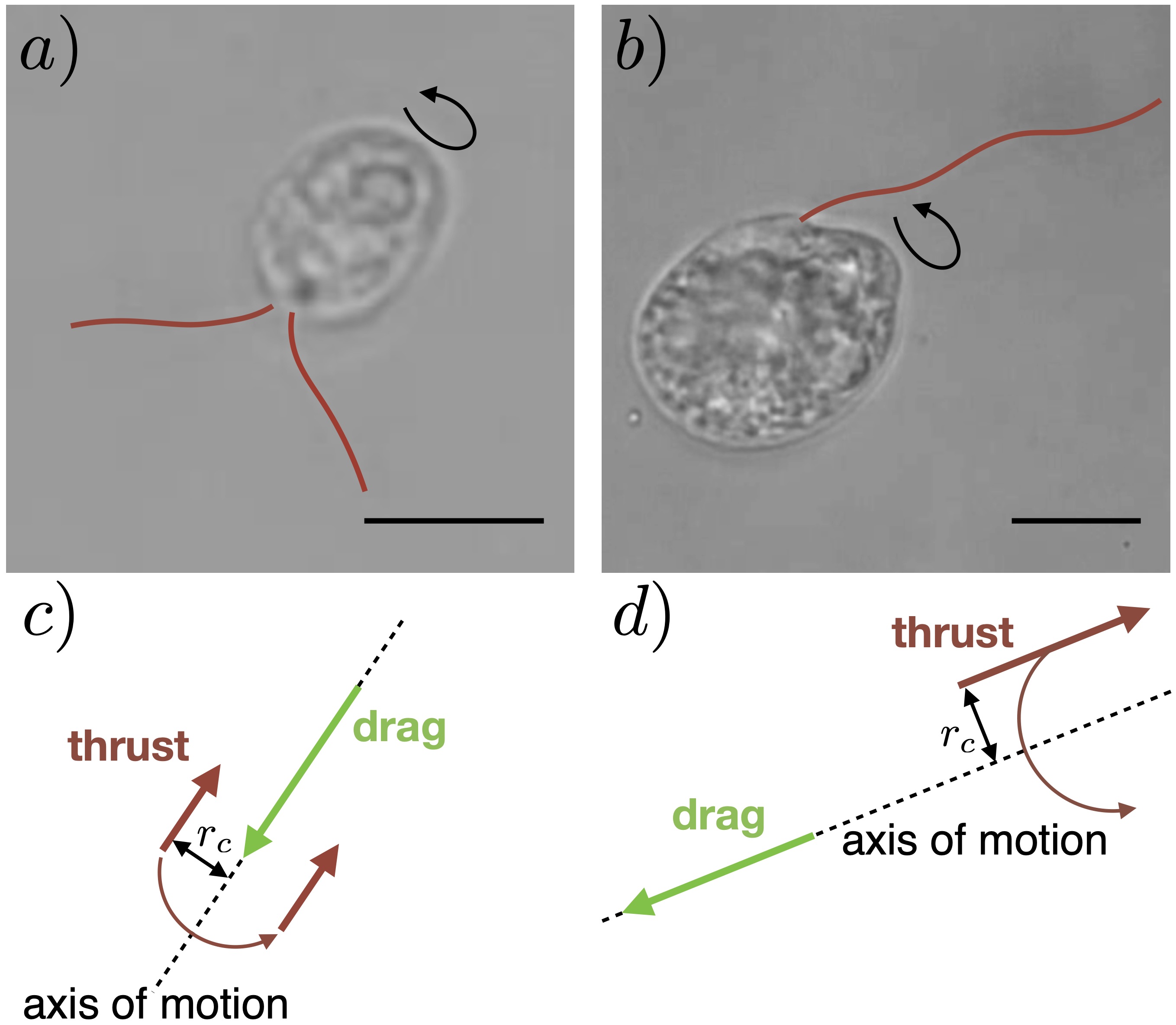}
\caption{a) and b) Photographs of \textit{Chlamydomonas reinhardtii} (wild-type) and \textit{Oxyrrhis marina} respectively. Flagella have been highlighted in red. These appendages are off-centered with respect to the body and rotates around the axis of motion as the cell spins. Scale bar represents $10 {\rm \mu m}$ in both cases. c) and d) To quantify the role of cell spinning on the experimental flow-fields in confinement, we have first modelled (see Fig.~\ref{figureS6}) these organisms as 3-forces (resp. 2-forces) swimmers with the Stokeslets approximation of Liron and Mochon \cite{Liron1976} and with the thrust force(s) that rotate(s) around the axis of motion in a circular orbit with radius $r_{c}$.}
\label{figureS2}
\end{wrapfigure}

\begin{wrapfigure}{r}{8cm}
\centering
\includegraphics[width=0.85\linewidth]{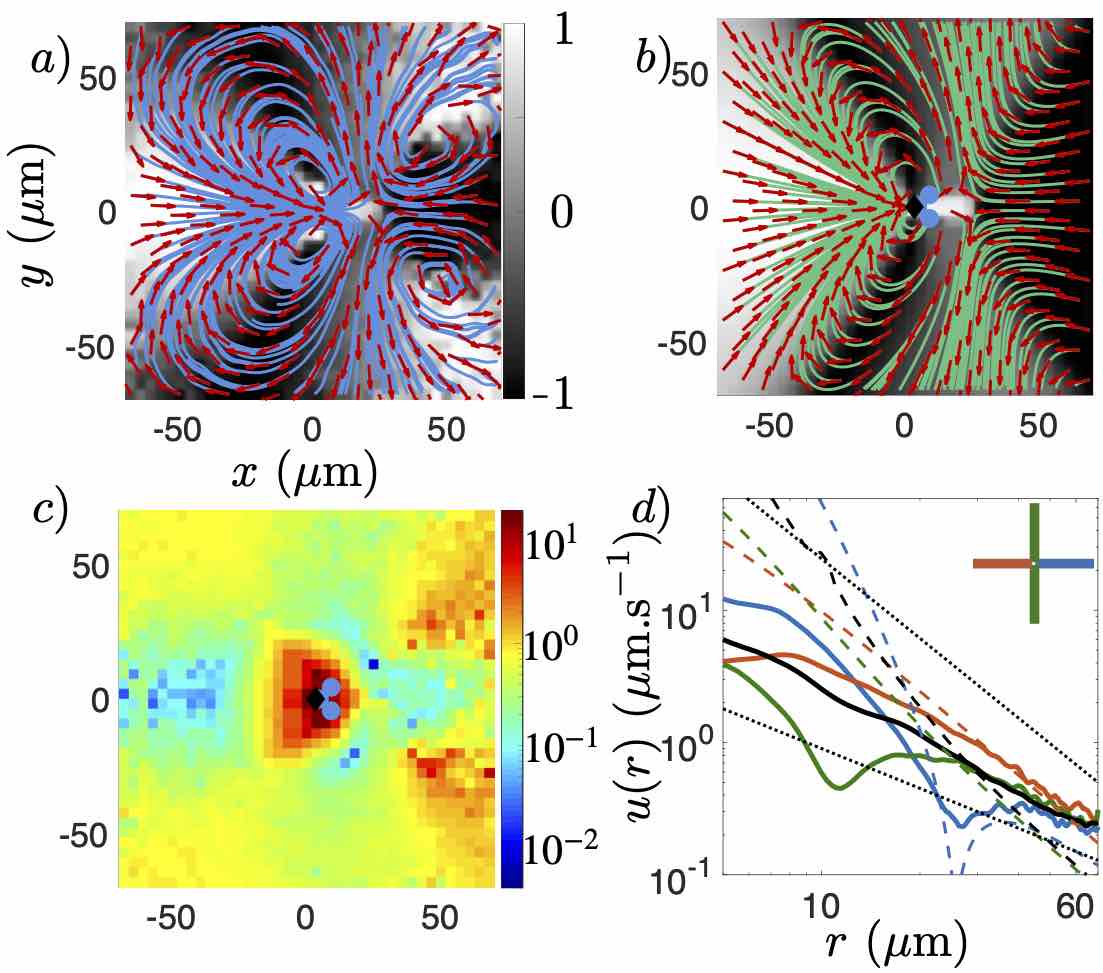}
\caption{Bulk flow fit. a) Experimental flow-field obtained for CR in $H=60 {\rm \mu m}$ (same as Fig. 2a main text). b) Best fit to a bulk model \cite{Drescher2010} obtained by considering a spinning 3-forces swimmer with the bulk Stokeslet solution and an additional 3D source dipole. The streamlines are qualitatively different. c) Relative error of the fit with respect to the experimental flow field. d) Log-log plot of the velocity decay of both the experimental (solid lines) and fitting flow-fields (dashed lines). Black dashed lines indicate $r^{-2}$ and $r^{-1}$ decay and are added as a guide to the eye.}
\label{figureS3}
\end{wrapfigure}

\begin{wrapfigure}{r}{8cm}
\centering
\includegraphics[width=0.95\linewidth]{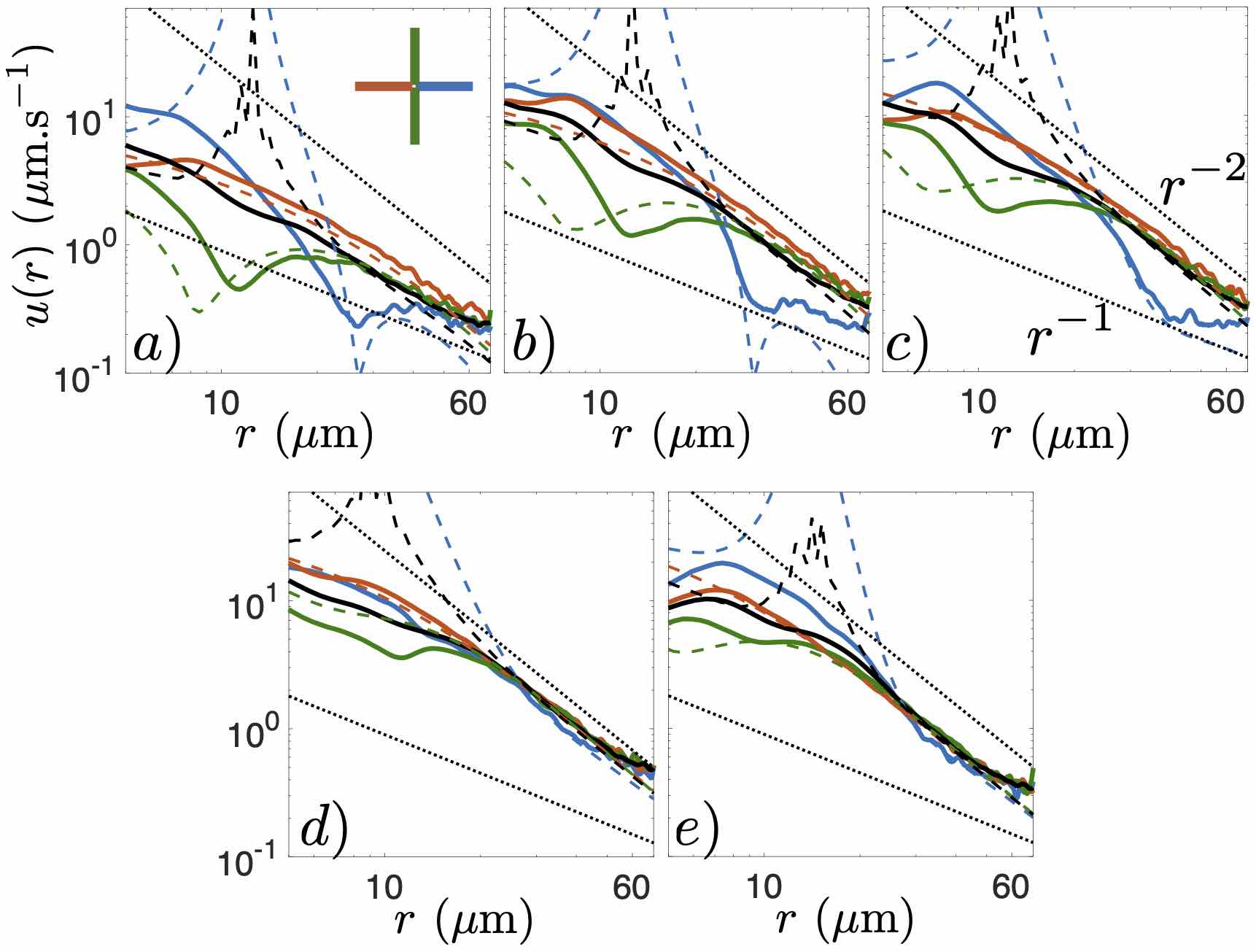}
\caption{Transects for the CR flow fields under different levels of confinement. Log-log plot of the velocity decay of both the experimental (solid lines) and fitting flow-fields (dashed lines) for: a) $H=60\,\mu$m, b) $H=43\,\mu$m, c) $H=29\,\mu$m, d) $H=21.5\,\mu$m, e) $H=14\,\mu$m. Black dashed lines indicate $r^{-2}$ and $r^{-1}$ decay and are added as a guide to the eye.}
\label{figureS4}
\end{wrapfigure}

\begin{wrapfigure}{r}{8cm}
\centering
\includegraphics[width=0.85\linewidth]{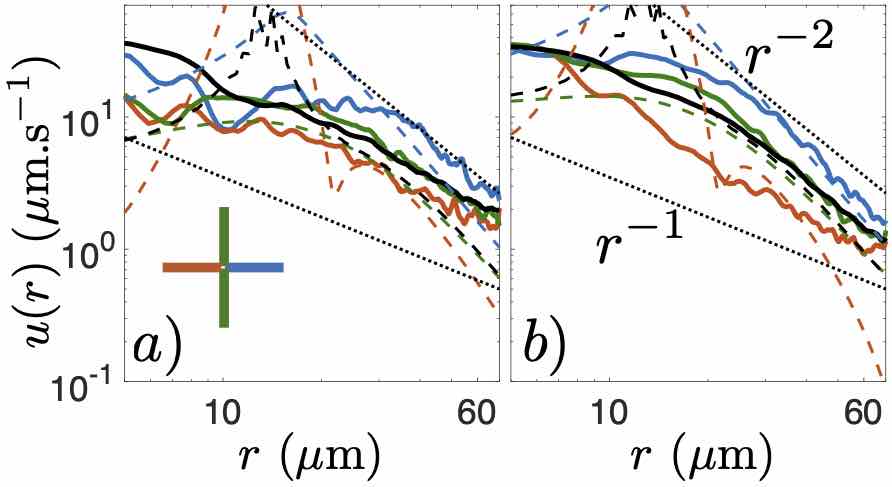}
\caption{Transects for the OM flow fields under different levels of confinement. Log-log plot of the velocity decay of both the experimental (solid lines) and fitting flow-fields (dashed lines) for: a) $H=29\,\mu$m, b) $H=21.5\,\mu$m. Black dashed lines indicate $r^{-2}$ and $r^{-1}$ decay and are added as a guide to the eye.}
\label{figureS5}
\end{wrapfigure}

\begin{wrapfigure}{r}{8cm}
\centering
\includegraphics[width=0.85\linewidth]{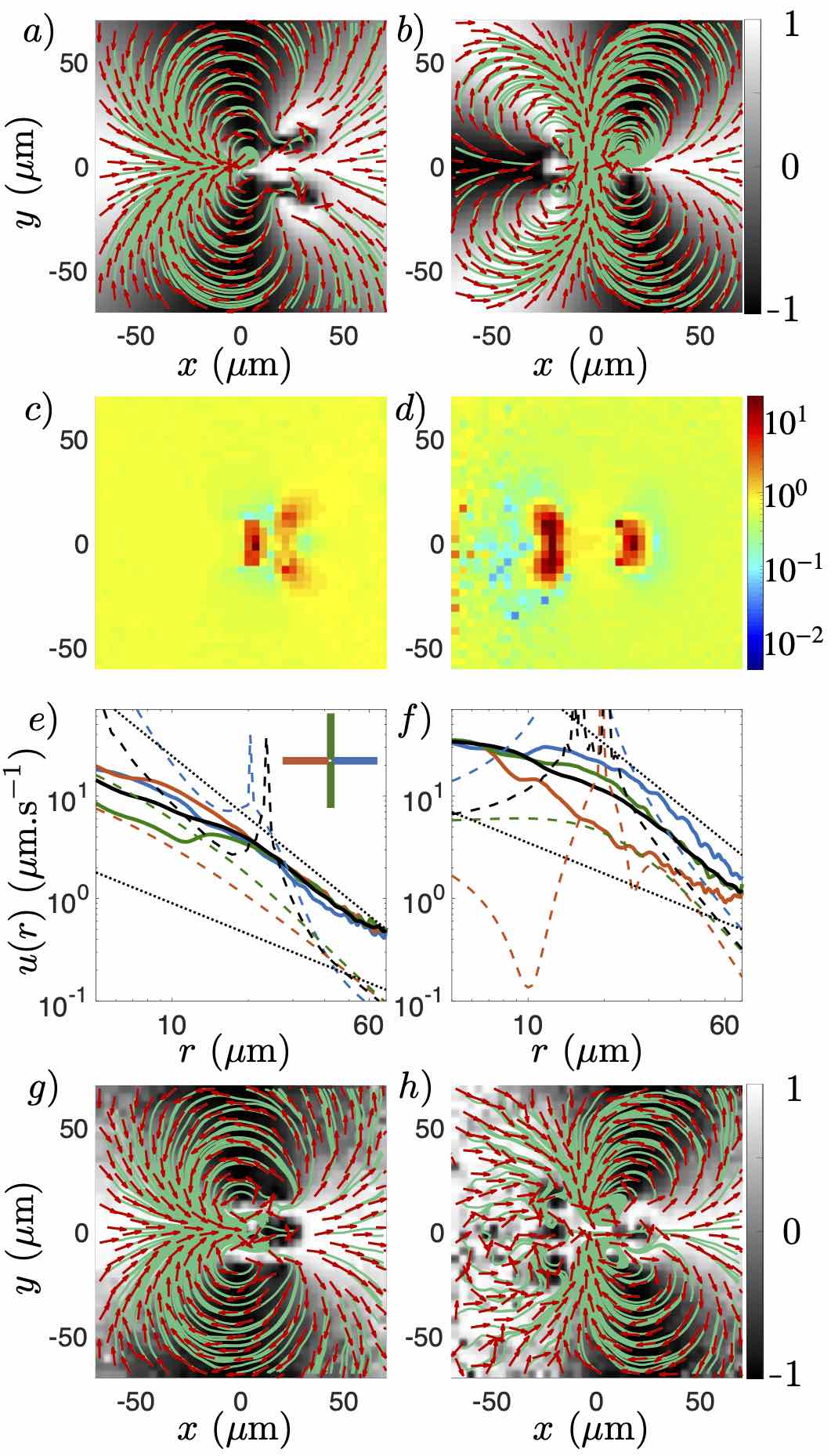}
\caption{Best fitting flow-fields obtained from the Stokeslet approximation in confinement of Liron and Mochon \cite{Liron1976} for a) CR and b) OM in $H=21.5 {\rm \mu m}$. The angular symmetry of the flow is well captured (compare panel a with Fig. 2b main text and panel b with Fig. 3a main text). c,d) 
Relative error of the fit with respect to the experimental flow field for CR and OM respectively. This is everywhere at least $\mathcal{O}(1)$. e,f) Log-log plot of the velocity decay of both the experimental (solid lines) and fitting flow-fields (dashed lines). The model does not describe well the experiments. g,h) Substracting the best fit to the corresponding experimental flow-field reveals a (noisy) dipolar symmetry. This reflects the fact that this approach does not take into account the finite size of the organisms.}
\label{figureS6}
\end{wrapfigure}

\begin{wrapfigure}{r}{8cm}
\centering
\includegraphics[width=\linewidth]{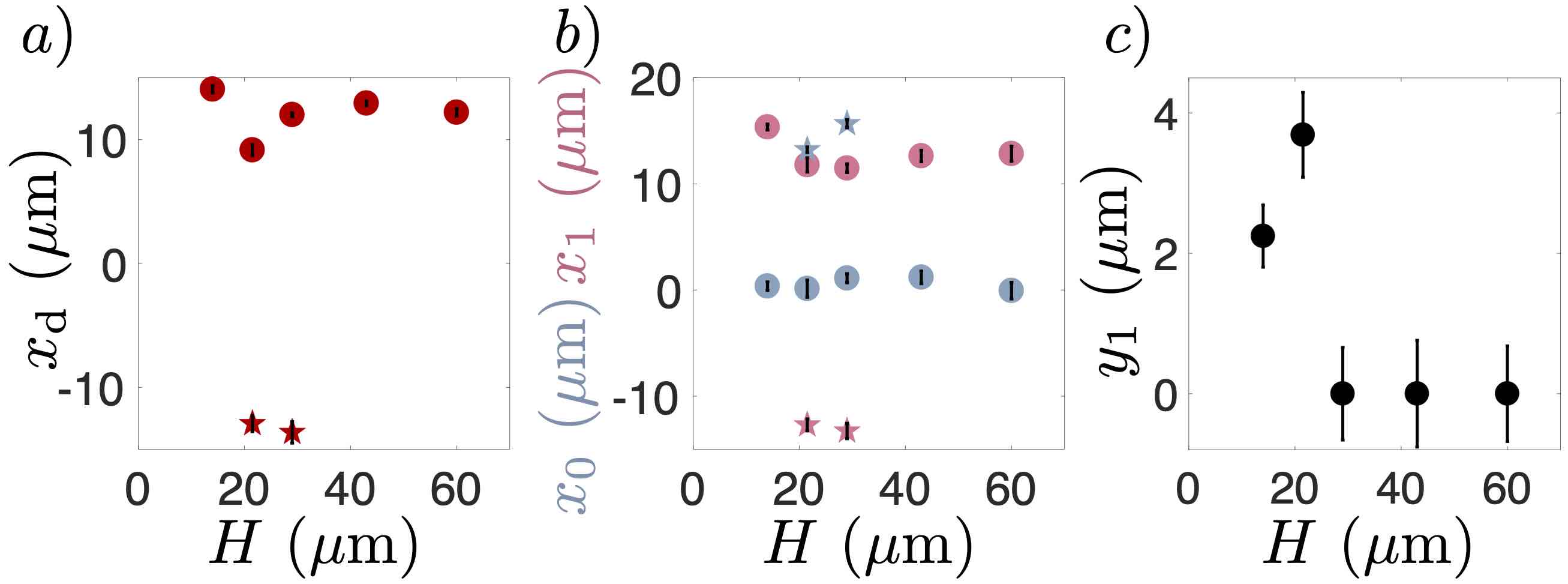}
\caption{Evolution of the other free-fitting parameters obtained from our effective 2D model. Stars = OM; circles = CR. a) Source dipole localisation $x_{\rm d}$. We consistently observe the dipole to be shifted towards the flagella (i.e. in front of the cell body center for CR and at the back for OM). b) On-axis position of the drag force $x_0$ (gray) and thrust force(s) $x_1$. c) Orthogonal position of the thrust forces $y_1$ in the case of CR.}
\label{figureS7}
\end{wrapfigure}

\begin{wrapfigure}{r}{8cm}
\centering
\includegraphics[width=0.6\linewidth]{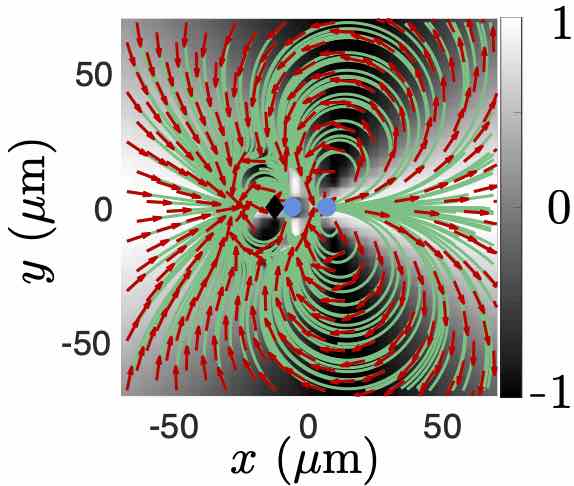}
\caption{Flow-field obtained from the best fitting parameters of OM's flow-field in $H=21.5 {\rm \mu m}$ with the exception that the distance $|x_0-x_1|$ between the 2 Stokeslets (blue circles) has been reduced by half. This flow-field has a dipolar symmetry already within the accessible experimental range as opposed to the measured flow-field (Fig. 3a-main text) showing that the spatial arrangement of the forces have a large impact in confinement.}
\label{figureS8}
\end{wrapfigure}

\end{document}